\documentclass[twocolumn,10pt]{article}    
\usepackage{url}  
\usepackage[utf8]{inputenc} 
\usepackage{CJKutf8}
\usepackage{graphicx}
\usepackage{authblk}

\begin{document}

\title{A genetic variant near olfactory receptor genes influences cilantro preference}
\author[1,*]{Nicholas Eriksson}
\author[1]{Shirley Wu}
\author[1]{Chuong B. Do}
\author[1]{Amy K. Kiefer}
\author[1]{Joyce Y. Tung}
\author[1]{Joanna L. Mountain}
\author[1]{David A. Hinds}
\author[1]{Uta Francke}
\affil[1]{23andMe, Inc., Mountain View, CA USA}
\affil[*]{\texttt{nick@23andme.com}}


\maketitle 

\begin{abstract} 
The leaves of the \textit{Coriandrum sativum} plant, known as cilantro or
coriander, are widely used in many cuisines around the world. However, far from
being a benign culinary herb, cilantro can be polarizing---many people love it
while others claim that it tastes or smells foul, often like soap or
dirt.  This soapy or pungent aroma is largely attributed to several aldehydes present in cilantro.
Cilantro preference is suspected to
have a genetic component, yet to date nothing is known about specific
mechanisms.
%
Here we present the results of a genome-wide association study 
among 14,604 participants of European ancestry who reported whether cilantro tasted
soapy, with replication  in a distinct set of 11,851 participants who declared whether they
liked cilantro.  We find a single nucleotide polymorphism (SNP) significantly
associated with  soapy-taste detection that is confirmed in the
cilantro preference group.  This SNP, rs72921001, ($p=6.4\cdot 10^{-9}$, odds
ratio 0.81 per A allele)  lies within a cluster of olfactory
receptor genes on chromosome 11.  Among these olfactory receptor genes is
\textit{OR6A2}, which has a high binding specificity for several of the
aldehydes that give cilantro its characteristic odor.
We also estimate the heritability of cilantro soapy-taste detection in our cohort, showing
that the heritability tagged by common SNPs is low, about 0.087.
%
These results confirm that there is a genetic component to cilantro taste
perception and suggest that cilantro dislike may stem from genetic variants in  olfactory
receptors.  We propose that \textit{OR6A2} may be the olfactory receptor that
contributes to the detection of a soapy smell from cilantro in European populations.
\end{abstract}

\section*{Background}
The \textit{Coriandrum sativum} plant has been cultivated since at least the
2nd millennium BCE \cite{Zohary-Hopf}. Its fruits (commonly called coriander
seeds) and leaves (called cilantro or coriander) are important components of
many cuisines. In particular, South Asian cuisines use both the leaves and the
seeds prominently, and Latin American food often incorporates the leaves.

The desirability of cilantro has been debated for centuries.  Pliny claimed
that coriander had important medicinal properties: \textit{``vis magna ad
refrigerandos ardores viridi''} (``While green, it is possessed of very cooling
and refreshing properties'') \cite{naturalhistory}.  The Romans used the leaves
and seeds in many dishes, including moretum (a herb, cheese, and garlic spread
similar to today's pesto) \cite{faas2002}; the Mandarin word for cilantro,
\begin{CJK}{UTF8}{min}香菜\end{CJK} (xi\={a}ngc\`{a}i), literally means
``fragrant greens''.  However, the leaves in particular have long inspired
passionate hatred as well; e.g., John Gerard called it a ``very stinking
herbe'' with leaves of ``venemous quality'' \cite{gerard1597, leach2001}.

It is not known why cilantro is so differentially perceived.  The proportion of
people who dislike cilantro varies widely by ancestry \cite{mauer2012};
however, it is not clear to what extent this may be explained by differences in
environmental factors, such as frequency of exposure.  Genetics has been
thought to play a role, but to date no studies have found genetic variants
influencing cilantro taste preference.

The smell of cilantro is often described as pungent or soapy.  It is suspected,
although not proven, that cilantro dislike is largely driven by the odor rather
than the taste.  The key aroma components in cilantro consist of various
aldehydes, in particular (E)-2-alkenals and n-aldehydes \cite{cadwallader2005,
pmid16013833}.  The unsaturated aldehydes (mostly decanal and dodecanal) in
cilantro are described as fruity, green, and pungent; the (E)-2-alkenals
(mostly (E)-2-decenal and (E)-2-dodecenal) as soapy, fatty, ``like cilantro'',
or pungent \cite{cadwallader2005, pmid16013833}.

Several families of genes are important for taste and smell. The TAS1R and
TAS2R families form sweet, umami, and bitter taste receptors
\cite{pmid11917125,pmid10761935}. The olfactory receptor family contains about
400 functional genes in the human genome.  Each receptor binds to a set of
chemicals, enabling one to recognize specific odorants or tastants.  Genetic
differences in many of these receptors are known to play a role in how we
perceive tastes and smells \cite{pmid21036327, pmid12595690, pmid17873857,
pmid20585627}. 

\section*{Results and discussion}

\begin{figure*}
\centering
\includegraphics[width=\textwidth]{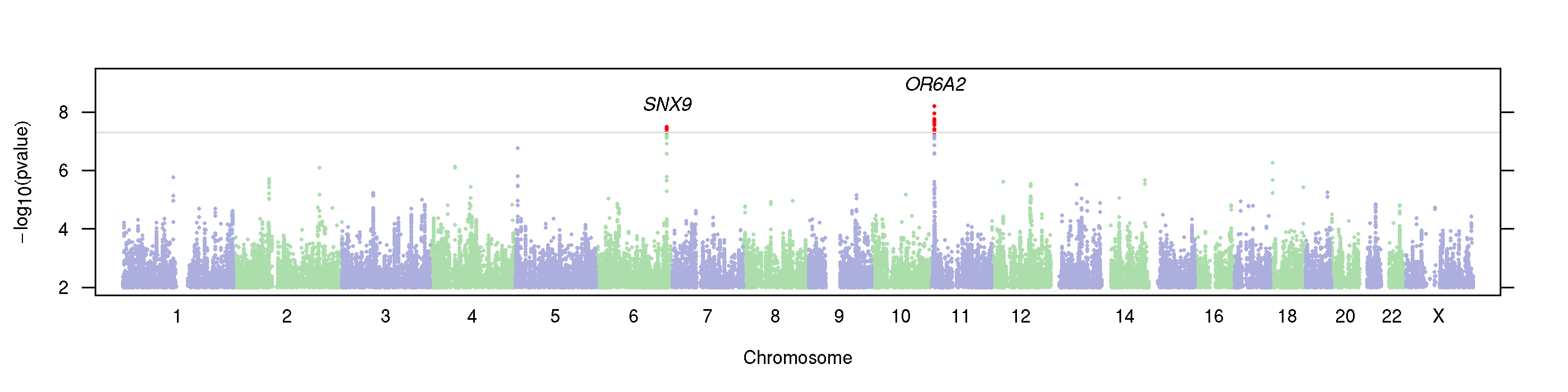}
\caption{\textbf{Manhattan plot of association with cilantro soapy-taste}
Negative $\log_{10} p$-values across all SNPs tested. SNPs shown in red are genome-wide
significant ($p < 5 \cdot 10^{-8}$).  Regions are named with the postulated
candidate gene.}
\label{fig:manhattan}
\end{figure*}

\begin{figure}
\centering
\includegraphics[width=.35\textwidth]{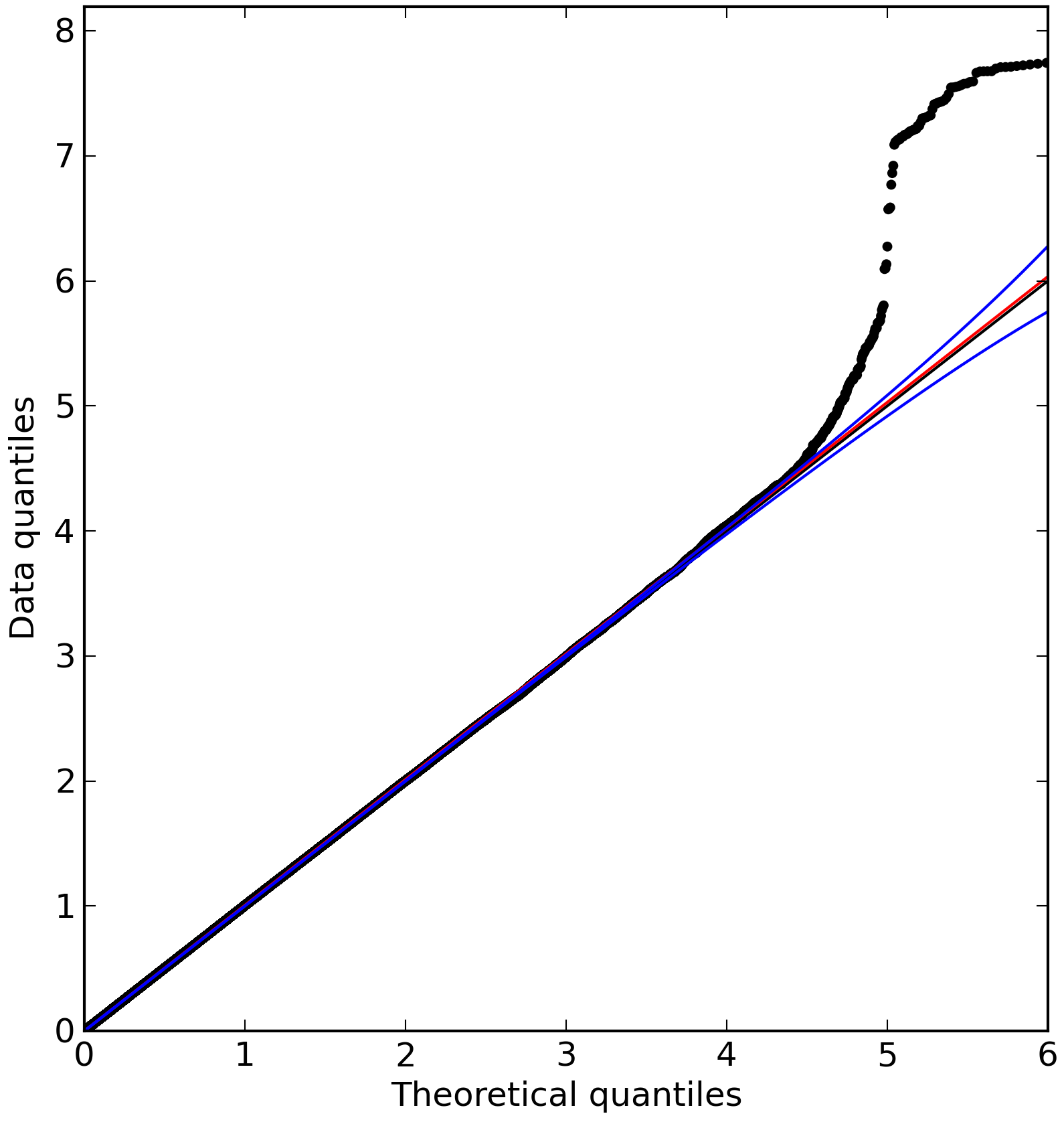}
\caption{\textbf{Quantile-quantile plot of association with cilantro soapy-taste}
Observed $p$-values versus theoretical $p$-values under the null hypothesis of
no association.  The genomic control inflation factor for the study was 1.007
and is indicated by the red line; approximate 95\% confidence intervals are
given by the blue curves.  } 
\label{fig:qq}
\end{figure}

\begin{figure}
\centering
\includegraphics[width=.5\textwidth]{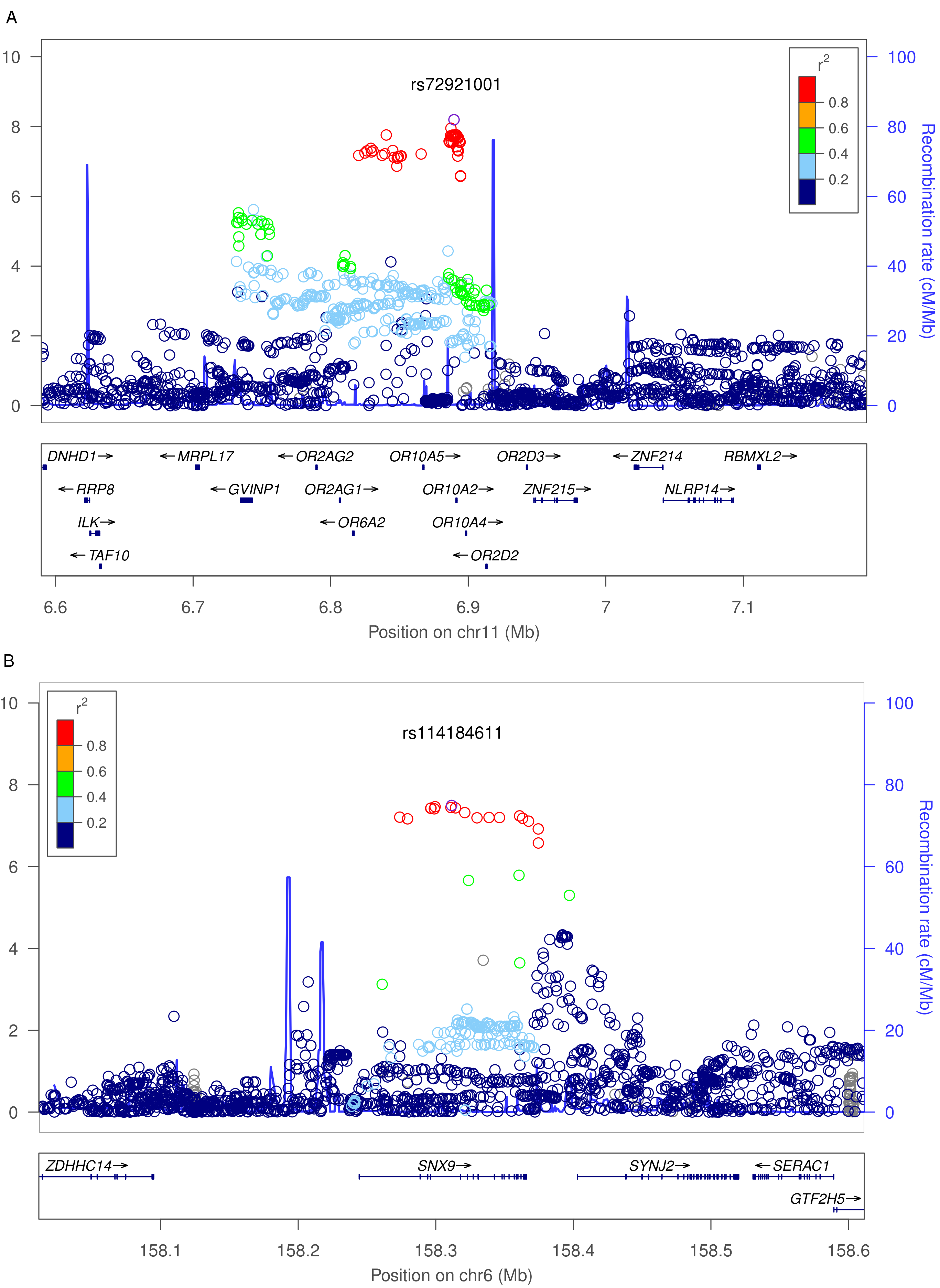}
\caption{\textbf{Associations with cilantro soapy-taste near rs72921001 (A) and rs114184611 (B)} 
Colors depict the squared correlation ($r^2$) of each SNP with the most
associated SNP (rs72921001, shown in purple). Gray indicates SNPs for which
$r^2$ information was missing.  }
\label{fig:region}
\end{figure}

Here we report the first ever genome-wide association study (GWAS) of cilantro
soapy-taste detection.   Briefly, the GWAS
was conducted in 14,604 unrelated participants of primarily European ancestry
who responded to an online questionnaire asking whether they thought cilantro
tasted like soap (Table~\ref{tab:cohort}). Two single nucleotide polymorphisms (SNPs) were significant genome-wide  ($p < 5\cdot
10^{-8}$) in this population.  One SNP, in a cluster of olfactory receptors,
replicated in a non-overlapping group of 11,851 participants (again, unrelated
and of primarily European ancestry) who reported whether they liked or disliked
cilantro  (see Methods for full details). Figure~\ref{fig:manhattan} shows $p$-values across the
whole genome; Figure~\ref{fig:region} shows $p$-values near the most significant associations.
A quantile-quantile plot (Figure~\ref{fig:qq}) shows little ($\lambda=1.007$)
global inflation of $p$-values.  Index SNPs with $p$-values under $10^{-6}$ are
shown in Table~\ref{tab:snps} (along with replication $p$-values). 

\begin{table}
\centering
\begin{tabular}{cccc}
\hline
\hline
        & N & Female & Age (SD) \\ 
\hline
Tastes soapy        & 1994 & 0.566 & 49.0 (15.0) \\
Doesn't taste soapy &12610 & 0.489 & 48.3 (15.2) \\
Total               &14604 & 0.500 & 48.4 (15.2) \\
\hline
Dislikes cilantro   &3181  & 0.487 & 47.1 (16.6) \\
Likes    cilantro   &8906  & 0.420 & 43.8 (14.5) \\
Total               &12087 & 0.438 & 44.7 (15.1) \\
\hline 
\hline
\end{tabular} 
\caption{Summary of the cohorts used in the analysis}
\label{tab:cohort}
\end{table}

\begin{table*}
\centering
\scriptsize
\begin{tabular}{cccccccccc}
\hline\hline
           SNP &chr&        pos&                    gene &allele &   MAF & $r^2$& $p_{\rm{discovery}}$ &  $p_{\rm{repl}}$ & OR (CI)\\ 
 \hline
  rs72921001& 11 &   6,889,648&            \textit{OR6A2} & C/A& 0.364 &0.969 & $6.4\cdot 10^{-9}$ & 0.0057 & 0.809 (0.753 -- 0.870)\\
 rs114184611&  6 & 158,311,499&             \textit{SNX9} & C/T& 0.077 &0.980 & $3.2\cdot 10^{-8}$ &   0.49 & 0.679 (0.588 -- 0.784)\\
chr5:4883483&  5 &   4,883,483&         \textit{ADAMTS16} & C/T& 0.032 &0.885 & $1.7\cdot 10^{-7}$ &   0.51 & 0.526 (0.405 -- 0.683)\\
   rs7227945& 18 &   4,251,279& \textit{DLGAP1/LOC642597} & T/G& 0.055 &0.920 & $5.3\cdot 10^{-7}$ &   0.96 & 1.447 (1.258 -- 1.663)\\
   rs6554267&  4 &  56,158,891&       \textit{KDR/SRD5A3} & T/G& 0.019 &0.651 & $7.4\cdot 10^{-7}$ &   0.85 & 1.975 (1.529 -- 2.549)\\
  rs13412810&  2 & 192,420,461&     \textit{MYO1B/OBFC2A} & G/A& 0.141 &0.942 & $7.9\cdot 10^{-7}$ &   0.78 & 0.770 (0.693 -- 0.857)\\
\hline\hline
\end{tabular} 
\caption{\textbf{Index SNPs for regions with $p < 10^{-6}$ for cilantro-soapy taste}
The index SNP is defined as the SNP with the smallest $p$-value within a
region.  The listed gene is our postulated candidate gene near the SNP.
Alleles are listed as major/minor (in Europeans). MAF is the frequency of the
minor allele in Europeans and $r^2$ is the estimated imputation accuracy.
$p_{\rm{discovery}}$ and $p_{\rm{repl}}$ are the discovery and replication
$p$-values, respectively.  The OR is the discovery odds ratio per copy of the
minor allele (e.g., the A allele of rs72921001 is the allele associated with a
lower risk of detecting a soapy taste).  }
\label{tab:snps}
\end{table*}

\begin{table*}
\centering
\begin{tabular}{lrrrcc}
\hline \hline
Population      & Not soapy (\%) & Soapy (\%)    & Total  &  MAF   & $p$-value\\
\hline
Ashkenazi       &   634 (85.9\%) &  104 (14.1\%) &   738  &  0.355 & 0.56 \\
South Europe    &   458 (86.6\%) &   71 (13.4\%) &   529  &  0.335 & 0.25 \\
Europe all      & 13213 (87.0\%) & 1973 (13.0\%) & 15186  &  0.373 & $1.23\cdot10^{-8}$\\
North Europe    & 11794 (87.2\%) & 1736 (12.8\%) & 13530  &  0.376 & $1.17\cdot10^{-8}$\\
All             & 16196 (87.6\%) & 2299 (12.4\%) & 18495  &  0.356 & $3.94\cdot10^{-8}$\\
African-American&   545 (90.8\%) &   55  (9.2\%) &   600  &  0.224 & 0.87\\
Latino          &   820 (91.3\%) &   78  (8.7\%) &   898  &  0.350 & 0.29\\
East Asia       &   424 (91.6\%) &   39  (8.4\%) &   463  &  0.283 & 0.22\\
South Asia      &   322 (96.1\%) &   13  (3.9\%) &   335  &  0.371 & 0.0078\\
\hline \hline
\end{tabular}
\caption{\textbf{Cilantro soapy-taste by ancestry} 
Number of people detecting a soapy taste by ancestry group, sorted from most
to least soapy-taste detection.  For reference, we have added the minor allele
frequency of rs7107418 in each group.  This SNP is a proxy for rs72921001 ($r^2
> .98$), with the minor G allele of rs7107418 corresponding to the minor A
allele of rs72921001 (which is associated with less soapy tasting). The $p$-value is the
$p$-value of association between soapy-taste and rs7107418 in each group.  }
\label{tab:ethnic}
\end{table*}

We found one significant association for cilantro soapy-taste that was
confirmed in the cilantro preference population.  The SNP,  rs72921001
($p_{\rm{discovery}}=6.4\cdot 10^{-9}$, OR=0.81, $p_{\rm{repl}} = 0.0057$) lies
on chromosome 11 within a cluster of eight olfactory receptor genes:
\textit{OR2AG2}, \textit{OR2AG1}, \textit{OR6A2}, \textit{OR10A5},
\textit{OR10A2}, \textit{OR10A4}, \textit{OR2D2}, and \textit{OR2D3}.  The C allele is associated 
with both detecting a soapy smell and disliking cilantro. Of the
olfactory receptors encoded in this region, \textit{OR6A2} appears to be the
most promising candidate underlying the association with cilantro odor
detection.  It is one of the most studied olfactory receptors  (often as the
homologous olfactory receptor I7 in rat) \cite{pmid20608641, pmid11100145,
pmid9875846, pmid14724183}.  A wide range of odorants have been found to
activate this receptor, all of them aldehydes \cite{pmid11100145}.  Among the
unsaturated aldehydes, octanal binds the best to rat I7 
\cite{pmid9875846}; however, compounds ranging from heptanal to
undecanal also bind to this receptor \cite{pmid11100145}.  Several singly
unsaturated n-aldehydes also show high affinity, including (E)-2-decenal
\cite{pmid11100145}.  These aldehydes include several of those playing a key
role in cilantro aroma, such as decanal and (E)-2-decenal.  Thus, this gene is
particularly interesting as a candidate for cilantro odor detection.  
The index SNP is also in high LD ($r^2 > 0.9$) with three non-synonymous SNPs
in \textit{OR10A2}, namely rs3930075, rs10839631, and rs7926083 (H43R, H207R, and K258T, respectively).
Thus \textit{OR10A2} may also be a reasonable candidate gene in this region.

The second significant association, with rs114184611 ($p_{\rm{discovery}}=3.2\cdot 10^{-8}$,
OR=0.68, $p_{\rm{repl}}= 0.49$), lies in an intron of the gene \textit{SNX9} (sorting nexin-9).  See Figure~\ref{fig:region}.
\textit{SNX9} encodes a multifunctional protein involved in intracellular
trafficking and membrane remodeling during endocytosis \cite{pmid19192055}. It
has no known function in taste or smell and did not show association with
liking cilantro in the replication population.
This SNP is located about 80kb upstream of \textit{SYNJ2}, an inositol
5-phosphatase thought to be involved in membrane trafficking and signal
transduction pathways.  In candidate gene studies, \textit{SYNJ2} SNPs were
found to be associated with agreeableness and symptoms of depression in the
elderly \cite{pmid22213687} and with cognitive abilities \cite{pmid22045296}.
In mice, a \textit{Synj2} mutation causes recessive non-syndromic hearing loss \cite{pmid21423608}.
Given recent evidence that the perception of flavor may be influenced by
multiple sensory inputs (cf.\ \cite{pmid22717401,pmid22717402}) we cannot exclude the
\textit{SYNJ2}-linked SNP as conveying a biologically meaningful association.
While this SNP may be a false positive,  it could also be the case that this SNP is associated only with detecting a
soapy smell in cilantro (and not in liking cilantro).

We have used two slightly different phenotypes in our discovery and replication,  soapy-taste
detection and cilantro preference, which are correlated ($r^2 \approx
0.33$). Detection of a soapy taste is reportedly one of the major reasons people
seem to dislike cilantro. Despite having over 10,000 more people reporting
cilantro preference, we have used soapy-taste detection as our primary
phenotype because it is probably influenced by fewer environmental factors.
Indeed, we see a stronger effect of rs72921001 on soapy-taste detection
than on cilantro preference (OR of 0.81 versus 0.92).  

We find significant differences  by sex and ancestral population in soapy-taste
detection (Tables~\ref{tab:cohort} and \ref{tab:ethnic}).  Women are more likely to detect a soapy taste (and
to dislike cilantro) (OR for soapy-taste detection 1.36, $p=2.5\cdot
10^{-10}$), Table~\ref{tab:cohort}.  African-Americans, Latinos, East Asians, and South Asians
are all significantly less likely to detect a soapy taste compared to
Europeans (ORs of 0.676, 0.637, 0.615, and 0.270 respectively, $p<0.003$), see
Table~\ref{tab:ethnic}. Ashkenazi Jews and South Europeans did not show significant
differences from Northern Europeans ($p= 0.84, 0.65$ respectively).
We tested the association between rs72921001 and soapy-taste detection within each population.
Aside from the European populations, there was only a significant association
in the small South Asian group ($p=0.0078$, OR=0.18, 95\% CI 0.053--0.64). This
association is in the same direction as the association in Europeans.  Note
that the GWAS population in Table~\ref{tab:cohort} is a subset of the ``Europe all''
population in Table~\ref{tab:ethnic}, filtered to remove
relatives (Methods).
While the differences in allele frequency across populations do not explain 
the differences in soapy-taste detection, our analysis does suggest that this SNP
may affect soapy-taste detection in non-European populations as well.

We calculated the heritability for cilantro soapy-taste detection 
using the GCTA software \cite{pmid21167468}.
We found a low heritability  of 0.087 ($p=0.08$, 95\% CI (-0.037 -- 0.211)).
This estimate is a lower bound for the true heritability, as our estimate only takes into account 
heritability due to SNPs genotyped in this study.
While this calculation does not exclude a heritability of zero, 
the existence of the association with rs72921001 
does give a non-zero lower bound on the heritability.
Despite the strength of the association of the SNP near \textit{OR6A2}, it
explains only about 0.5\% of the variance in
perceiving that cilantro tastes soapy.  

There are a few possible explanations for these heritability numbers.  It is
possible that other genetic factors not detected here could influence cilantro
preference.  For example, there could be rare variants not typed in this study
(possibly in partial linkage disequilibrium with rs72921001) that have a larger
effect on cilantro preference.  Such rare variants could cause the true
heritability of this phenotype to be larger than we have calculated. For
example, the heritability of height is estimated to be about 0.8 
however, the heritability tagged by common SNPs is calculated at
about 0.45 \cite{pmid20562875}.
On the other hand, there is still considerable room between the 0.5\% variance
explained by  rs72921001 and the estimated heritability of 8.7\%. Thus it is
quite possible that cilantro preference could be polygenic, as many other
complex traits are (e.g., \cite{pmid20881960}).  
Finally, it is possible that the heritability of cilantro preference is
just rather low and that, aside from the association discovered here, there
is not a strong genetic component to cilantro preference.  We note that there
can be epigenetic modifiers of taste as well, for example, food preferences can
even be transmitted to the fetus in utero through the mother's diet
\cite{pmid22717401}.

\section*{Conclusions}

Through a GWAS, we have shown that a SNP, rs72921001, near a cluster of
olfactory receptors is significantly associated with detecting a soapy
taste to cilantro.  One of the  genes near this SNP encodes an olfactory receptor, OR6A2,
that detects the aldehydes that may make cilantro smell soapy and thus is a
compelling candidate gene for the  detection of the cilantro odors that give cilantro its
divisive flavor.

\section*{Methods}
\subsection*{Subjects}
Participants were drawn from the customer base of 23andMe, Inc., a consumer
genetics company. This cohort has been described in detail previously
\cite{pmid20585627, pmid21858135}.  Participants provided informed consent and
participated in the research online, under a protocol approved by the external
AAHRPP-accredited IRB, Ethical and Independent Review Services (E\&I Review).  

\subsection*{Phenotype data collection}
On the 23andMe website, participants contribute information through a
combination of research surveys (longer, more formal questionnaires) and
research ``snippets'' (multiple-choice questions appearing as part of various
23andMe webpages).  In this study, participants were asked two questions about
cilantro via research snippets:
\begin{itemize}
\item ``Does fresh cilantro taste like soap to you?'' (Yes/No/I'm not sure)
\item ``Do you like the taste of fresh (not dried) cilantro?'' (Yes/No/I'm not sure)
\end{itemize}
Among all 23andMe customers, 18,495 answered the first question (as either yes or no), 29,704 the second, and
15,751 both.  Participants also reported their age. Sex and ancestry were
determined on the basis of their genetic data.  From these answers, we chose a
set of 14,604 participants who answered the ``soapy'' question for GWAS, and
11,851 who answered only the taste preference question for a replication set.

In both the GWAS set and the replication set, all participants were of European
ancestry. In either group, no two shared more than 700 cM of DNA identical by
descent (IBD, approximately the lower end of sharing between a pair of first
cousins).  IBD was calculated using the methods described in
\cite{pmid22509285}; the principal component analysis was performed as in
\cite{pmid20585627}.  To determine European and African-American ancestry, we
used local-ancestry methods (as in \cite{pmid22493691}). Europeans had over 97\% of their genome painted European,
African-Americans had at least 10\% African and at most 10\% Asian ancestry.
Other groups were
built using anecstry informative markers trained on a subset of 23andMe
customers who reported having four grandparents of a given ancestry.

\subsection*{Genotyping}
Subjects were genotyped on one or more of three chips, two based on the
Illumina HumanHap550+ BeadChip, the third based on the Illumina OmniExpress+
BeadChip.  The  platforms contained 586,916, 584,942, and 1,008,948 SNPs.
Totals of 291, 5,394, and 10,184 participants (for the GWAS population) were
genotyped on the platforms, respectively. A total of 1,265 individuals were
genotyped on multiple chips.  For all participants, we imputed genotypes in batches of 8,000--10,000 using
Beagle and Minimac \cite{pmid17924348,pmid22820512,minimac} against the August 2010 release
of the 1000 Genomes reference haplotypes \cite{pmid20981092}, as described in
\cite{pmid22747683}.

A total of 11,914,767 SNPs were imputed.  Of these, 7,356,559  met our
thresholds of $0.001$ minor allele frequency, average $r^2$ across batches of
at least 0.5, and minimum $r^2$ across batches of at least 0.3.  The minimum
$r^2$ requirement was added to filter out SNPs that imputed less well in the
batches consisting of the less dense platform.  
Positions and alleles are given relative to the positive strand of build 37 of
the human genome.  

\subsection*{Statistical analysis}

For the GWAS, $p$-values were calculated using a likelihood ratio test for the
genotype term in the logistic regression model \[ Y \sim G + age + sex + pc_1 +
pc_2 + pc_3 + pc_4 + pc_5, \] where $Y$ is the vector of phenotypes (coded as
1=thinks cilantro tastes soapy, 0=doesn't), $G$ is the vector of genotypes 
(coded as a dosage 0--2 for the estimated number of minor alleles
present), and $pc_1, \dots, pc_5$ are the projections onto the principal
components. The same model was used for the replication, with the phenotype coded
as 1=dislikes cilantro, 0=likes.
We used the standard cutoff for genome-wide significance of $5\cdot 10^{-8}$ to
correct for the multiple tests in the GWAS.
ORs and $p$-values for the differences in soapy-taste detection between sexes and population
were calculated directly, without any covariates.  Table~\ref{tab:ethnic} uses a proxy SNP for rs72921001,
as our imputation was done only in Europeans, so we did not have data for rs72921001 in other populations.

For the heritability calculations, we used the GCTA software \cite{pmid21167468}.  The
calculations were done on genotyped SNPs only within a group of 13,628
unrelated Europeans.  Unrelated filtering here was done using GCTA to remove individuals
with estimated relatedness larger than 0.025. Thus, this group is slightly different from
the GWAS set, as there relatedness filtering there was done using IBD.
We assumed a
prevalence for soapy-taste detection
of 0.13 for the transformation of heritability from the 0-1 scale to the
liability scale. Otherwise, default options were used.  We calculated
heritability for autosomal and X chromosome SNPs separately; the estimates were
0.0869 (standard error 0.0634, $p$-value 0.0805) for autosomal SNPs and $2\cdot
10^{-6}$ (standard error 0.010753, $p$-value 0.5) for the X chromosome.

\section*{Author's contributions}
NE, SW, CBD, AKK, JLM, DAH, UF, and JYT conceived and designed the experiments.
NE analyzed the data and drafted the manuscript with contributions from all
other authors.

\section*{Acknowledgements}
We thank the customers of 23andMe for participating in this research and all
the employees of 23andMe for contributing to the research.


\end{document}